# Speaker Identification in a Shouted Talking Environment Based on Novel Third-Order Circular Suprasegmental Hidden Markov Models

Ismail Shahin

Department of Electrical and Computer Engineering

University of Sharjah

P. O. Box  27272

Sharjah, United Arab Emirates

Tel: (971) 6 5050967

Fax: (971) 6 5050877

E-mail: ismail@sharjah.ac.ae

**Abstract**

It is well known that speaker identification yields very high performance in a neutral talking environment; on the other hand, the performance has been sharply declined in a shouted talking environment. This work aims at proposing, implementing, and evaluating novel Third-Order Circular Suprasegmental Hidden Markov Models (CSPHMM3s) to improve the low performance of text-independent speaker identification in a shouted talking environment. CSPHMM3s possess combined characteristics of: Circular Hidden Markov Models (CHMMs), Third-Order Hidden Markov Models (HMM3s), and Suprasegmental Hidden Markov Models (SPHMMs). Our results show that CSPHMM3s are superior to each of: First-Order Left-to-Right Suprasegmental Hidden Markov Models (LTRSPHMM1s), Second-Order Left-to-Right Suprasegmental Hidden Markov Models (LTRSPHMM2s), Third-Order Left-to-Right Suprasegmental Hidden Markov Models (LTRSPHMM3s), First-Order Circular Suprasegmental Hidden Markov Models (CSPHMM1s), and Second-Order Circular Suprasegmental Hidden Markov Models (CSPHMM2s) in a shouted talking environment. Using our collected speech database, average speaker identification performance in a shouted talking environment based on LTRSPHMM1s, LTRSPHMM2s, LTRSPHMM3s, CSPHMM1s, CSPHMM2s, and CSPHMM3s is 74.6%, 78.4%, 81.7%, 78.7%, 83.4%, and 85.8%, respectively. Speaker identification performance that has been achieved based on CSPHMM3s is close to that attained based on subjective assessment by human listeners.

## 1. Introduction and Literature Review

The process of determining from which of the registered speaker a given utterance comes is defined as speaker identification. Speaker identification can be used in criminal investigations to determine the suspected persons produced the voice recorded at the scene of the crime. Speaker identification can also be used in civil cases or for the media. These cases include calls to radio stations, local or other government authorities, insurance companies, recorded conversations, and many other applications [1].

Speaker identification typically functions in one of two cases: text-dependent (fixed text) case or text-independent (free-text) case. In the text-dependent case, utterances of the same text are used for both training and testing. On the other hand, in the text-independent case, training and testing involve utterances from different texts. Speaker identification can be divided into two categories: "open set" and "closed set". In the "open set" category, a reference model for the unknown speaker may not exist; whereas, in the "closed set" category, a reference model for the unknown speaker should be available to the system.

Speaker identification performance in a neutral talking environment is extremely high [1-3]; however, the performance becomes very low in a stressful talking environment [4-11]. The talking environment in which speech is uttered assuming that speakers do not suffer from any stressful or emotional talking condition is called a neutral talking environment. The talking environment that makes speakers to vary their production of speech from a neutral talking condition to other stressful talking conditions such as shouted, loud, and fast is named a stressful talking environment.

Chen [4] studied talker-stress-induced intra-word variability and an algorithm that compensates for the systematic changes observed based on hidden Markov models (HMMs) trained by speech tokens under various talking conditions. Raja and Dandapat [5] studied speaker recognition under stressed conditions to enhance the declined performance under such conditions. Four different stressed conditions of SUSAS database [12], [13] have been used in their study. These conditions are neutral, angry, Lombard, and question. Their work [5] showed that the least speaker identification performance happened when speakers talk angrily. Angry talking environment has been used as an alternative talking environment to a shouted talking environment since it can not be completely separated from a shouted talking environment in our real life [6-10]. Zhang and Hansen [11] reported on the analysis of characteristics of speech in five different vocal modes: whispered, soft, neutral, loud, and shouted; and to identify categorizing features of speech modes.

In one of their works, Hadar and Messer [14] proposed an easy method based on transforming any high order HMM (including models in which the state sequence and observation dependency are of different orders) into an equivalent first order HMM. Chatzis [15] focused in one of his studies on proposing infinite-order HMMs to learn from data with sequential dynamics. These models typically depend on the postulation of first-order Markovian dependencies between the successive label values $y$. There are two main advantages of the designed models over the other approaches. The first advantage is that such models allow for capturing very long and complex temporal dependencies. The second advantage is that these models use a margin maximization paradigm to carry out model training, which yields a convex optimization design [15].

In five of our previous studies [6-10], we focused on improving poor speaker identification performance in a shouted talking environment based on different classifiers and models. Second-Order Hidden Markov Models (HMM2s) have been used to improve the recognition performance of isolated-word text-dependent speaker identification in a shouted talking environment [6]. The achieved speaker identification performance based on these models is 59.0%. Second-Order Circular Hidden Markov Models (CHMM2s) have been proposed, implemented, and tested to enhance the performance of isolated-word text-dependent speaker identification in such a talking environment [7]. Based on such models, the obtained speaker identification performance is 72.0%. In one of our works [8], we exploited Suprasegmental Hidden Markov Models (SPHMMs) to alleviate the degraded performance of text-dependent speaker identification in a shouted talking environment. The attained speaker identification performance using these models is 75.0%. Second-Order Circular Suprasegmental Hidden Markov Models (CSPHMM2s) have been proposed, applied, and evaluated to improve text-dependent speaker identification performance in a shouted talking environment [9]. The reported speaker identification performance based on these models is 83.4%. Novel Third-Order Hidden Markov Models (HMM3s) have been designed, implemented, and assessed to enhance the low performance of text-independent speaker identification in such a talking environment [10]. These novel models yield a text-independent speaker identification performance of 63.5%.

The main motivation of this research is to further enhance low text-independent speaker identification performance in a shouted talking environment over that obtained in the five prior works [6-10]. This will be achieved by proposing, implementing, and assessing novel classifiers called Third-Order Circular Suprasegmental Hidden Markov Models (CSPHMM3s). The proposed models are comprised of combinations from each of: CHMMs, SPHMMs, and HMM3s. We believe that CSPHMM3s will outperform each of: First-Order Left-to-Right Suprasegmental Hidden Markov Models (LTRSPHMM1s),

Second-Order Left-to-Right Suprasegmental Hidden Markov Models (LTRSPHMM2s), Third-Order Left-to-Right Suprasegmental Hidden Markov Models (LTRSPHMM3s), First-Order Circular Suprasegmental Hidden Markov Models (CSPHMM1s), and CSPHMM2s. This is because the characteristics of LTRSPHMM1s, LTRSPHMM2s, LTRSPHMM3s, CSPHMM1s, and CSPHMM2s are all combined and integrated into novel models called CSPHMM3s.

The remainder of the paper is structured as follows: The fundamentals of SPHMMs are given in Section 2. Section 3 summarizes LTRSPHMM1s, LTRSPHMM2s, CSPHMM1s, and CSPHMM2s. The details of CSPHMM3s are discussed in Section 4. Section 5 describes the collected speech database used in the experiments and the extraction of features. Section 6 discusses speaker identification algorithm and the experiments based on each of LTRSPHMM1s, LTRSPHMM2s, LTRSPHMM3s, CSPHMM1s, CSPHMM2s, and CSPHMM3s. Discussion of the achieved results appears in Section 7. Concluding remarks are given in Section 8.

## 2. Fundamentals of Suprasegmental Hidden Markov Models

Shahin has developed, implemented, and evaluated SPHMMs as classifiers for speaker recognition in stressful and emotional talking environments [8, 9, 16, 17] and for talking condition recognition in stressful and emotional talking environments [18, 19]. SPHMMs have proved to be superior classifiers over HMMs for speaker identification in each of shouted [8, 9] and emotional talking environments [16, 17] and for talking condition recognition in stressful and emotional talking environments [18, 19]. SPHMMs are capable of summarizing several states of HMMs into a new state called suprasegmental state. Suprasegmental state has the ability to look at the observation sequence through a larger window. This state permits observations at rates suitable for the situation of modeling. Prosodic information, as an example, can not be sensed at a rate that is used for acoustic modeling. The main acoustic parameters that describe prosody are fundamental frequency, intensity, and duration of speech signals [20]. Prosodic parameters of a unit of speech are named suprasegmental parameters since they have an impact on all segments of the unit of speech. Hence, prosodic events at the levels of phone, syllable, word, and utterance are expressed using suprasegmental states; on the other hand, acoustic events are expressed using conventional states.

The combination and integration of prosodic and acoustic information can be performed as given in the following formula [21],

$$log\ P(\lambda^v, \Psi^v | O) = (1-\alpha).\ log\ P(\lambda^v | O) + \alpha.\ log\ P(\Psi^v | O) \qquad (1)$$

where $\alpha$ is a weighting factor. When:

$$\begin{cases} 0.5 > \alpha > 0 & \text{biased towards acoustic model} \\ 1 > \alpha > 0.5 & \text{biased towards prosodic model} \\ \alpha = 0 & \text{biased completely towards acoustic model and} \\ & \text{no effect of prosodic model} \\ \alpha = 0.5 & \text{not biased towards any model} \\ \alpha = 1 & \text{biased completely towards prosodic model and} \\ & \text{no impact of acoustic model} \end{cases}$$

$\lambda^v$ is the acoustic model of the $v^{th}$ speaker, $\Psi^v$ is the suprasegmental model of the $v^{th}$ speaker, $O$ is the observation vector or sequence of an utterance, $P(\lambda^v | O)$ is the probability of the $v^{th}$ HMM speaker model given the observation vector $O$, and $P(\Psi^v | O)$ is the probability of the $v^{th}$ SPHMM speaker model given the observation vector $O$. More details about suprasegmental hidden Markov models can be obtained from references [8, 9, 16, 17, 18, 19].

## 3. Overview of: LTRSPHMM1s, LTRSPHMM2s, CSPHMM1s, and CSPHMM2s
### 3.1. First-order left-to-right suprasegmental hidden Markov models

LTRSPHMM1s have been derived from acoustic First-Order Left-to-Right Hidden Markov Models (LTRHMM1s). LTRHMM1s have been used in the last four decades in many works in the areas of speech, speaker, and emotion recognition since phonemes follow strictly left to right sequence [1, 3, 4, 22-26]. Fig. 1 illustrates an example of a basic structure of LTRSPHMM1s that has been obtained from LTRHMM1s. This figure illustrates an example of six first-order acoustic hidden Markov states ($q_1, q_2, ..., q_6$) with a left-to-right transition. In this figure, $p_1$ is a first-order suprasegmental state which is made up of $q_1, q_2$ and $q_3$, $p_2$ is a first-order suprasegmental state which is composed of $q_4, q_5$ and $q_6$. The suprasegmental states $p_1$ and $p_2$ are placed in a left-to-right form. $p_3$ is a first-order suprasegmental state which is comprised of $p_1$ and $p_2$. $a_{ij}$ is the transition probability between the $i^{th}$ and the $j^{th}$ acoustic hidden Markov states, while $b_{ij}$ is the transition probability between the $i^{th}$ and the $j^{th}$ suprasegmental states.

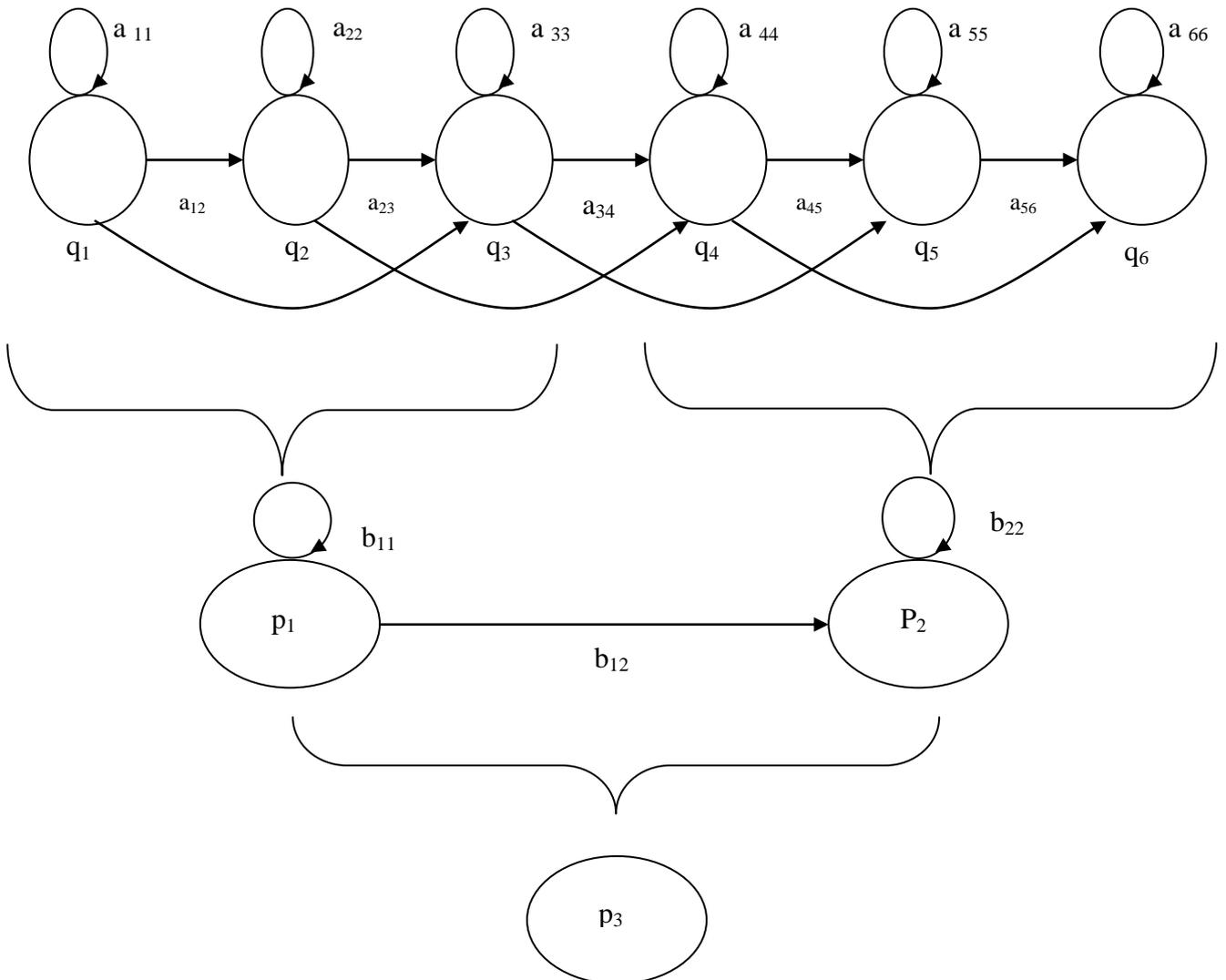

**Figure 1.** Basic structure of LTRSPHMM1s obtained from LTRHMM1s

In LTRHMM1s, the state sequence is a first-order Markov chain where the stochastic process is modeled in a 2-D matrix of a priori transition probabilities ($a_{ij}$) between states $s_i$ and $s_j$ where $a_{ij}$ are given as:

$$a_{ij} = \text{Prob}\left(q_t = s_j | q_{t-1} = s_i\right) \qquad (2)$$

In such acoustic models, it is assumed that the state-transition probability at time *t+1* depends only on the state of the Markov chain at time *t*. Readers can get more information about acoustic first-order left-to-right hidden Markov models from references [25, 26].

**3.2. Second-order left-to-right suprasegmental hidden Markov models**

LTRSPHMM2s have been derived from acoustic Second-Order Left-to-Right Hidden Markov Models (LTRHMM2s). As an example of such models, the six first-order acoustic left-to-right hidden Markov states of Fig. 1 have been changed by six second-order acoustic hidden Markov states positioned in a left-to-right form. The suprasegmental second-order states $p_1$ and $p_2$ are located in a left-to-right form. The suprasegmental state $p_3$ in such models becomes a second-order suprasegmental state.

In LTRHMM2s, the state sequence is a second-order Markov chain where the stochastic process is expressed by a 3-D matrix ($a_{ijk}$). Therefore, the transition probabilities in LTRHMM2s are given as [27]:

$$a_{ijk} = \text{Prob}\left(q_t = s_k | q_{t-1} = s_j, q_{t-2} = s_i\right) \qquad (3)$$

with the constraints,

$$\sum_{k=1}^{N} a_{ijk} = 1 \qquad N \geq i, j \geq 1$$

The state-transition probability in LTRHMM2s at time *t+1* relies on the states of the Markov chain at times *t* and *t-1*. More information about acoustic second-order left-to-right hidden Markov models can be obtained from references [6, 7, 27].

### 3.3. First-order circular suprasegmental hidden Markov models

CSPHMM1s have been obtained from acoustic First-Order Circular Hidden Markov Models (CHMM1s). CHMM1s have been proposed and implemented by Zheng and Yuan for speaker identification in a neutral talking environment [28]. Shahin showed that these models outperform LTRHMM1s for speaker identification in a shouted talking environment [7]. Interested readers can get more details about CHMM1s from references [7, 28].

Fig. 2 shows an example of a fundamental structure of CSPHMM1s that has been constructed from CHMM1s. This figure is made up of six first-order acoustic hidden Markov states: $q_1, q_2, ..., q_6$ arranged in a circular form. $p_1$ is a first-order suprasegmental state that is composed of $q_4, q_5$, and $q_6$. $p_2$ is a first-order suprasegmental state which consists of $q_1, q_2$, and $q_3$. The suprasegmental states $p_1$ and $p_2$ are placed in a circular form. $p_3$ is a first-order suprasegmental state that is comprised of $p_1$ and $p_2$.

### 3.4. Second-order circular suprasegmental hidden Markov models

CSPHMM2s have been derived from acoustic Second-Order Circular Hidden Markov Models (CHMM2s) [9]. CHMM2s have been proposed, applied, and evaluated by Shahin for speaker identification in each of shouted and emotional talking environments [7, 16]. CHMM2s have shown to be superior models over each of LTRHMM1s, LTRHMM2s, and CHMM1s because CHMM2s possess the characteristics of both CHMMs and HMM2s [7]. More information about CSPHMM2s can be obtained from reference [9].

### 4. Third-Order Circular Suprasegmental Hidden Markov Models

Third-Order Circular Suprasegmental Hidden Markov Models have been derived from acoustic Third-Order Hidden Markov Models (HMM3s). HMM3s have been proposed, implemented, and assessed by Shahin [10] to enhance the declined text-independent speaker identification performance in a shouted talking environment. In one of his works,

Shahin [10] showed that HMM3s outperform each of HMM1s and HMM2s in such a talking environment.

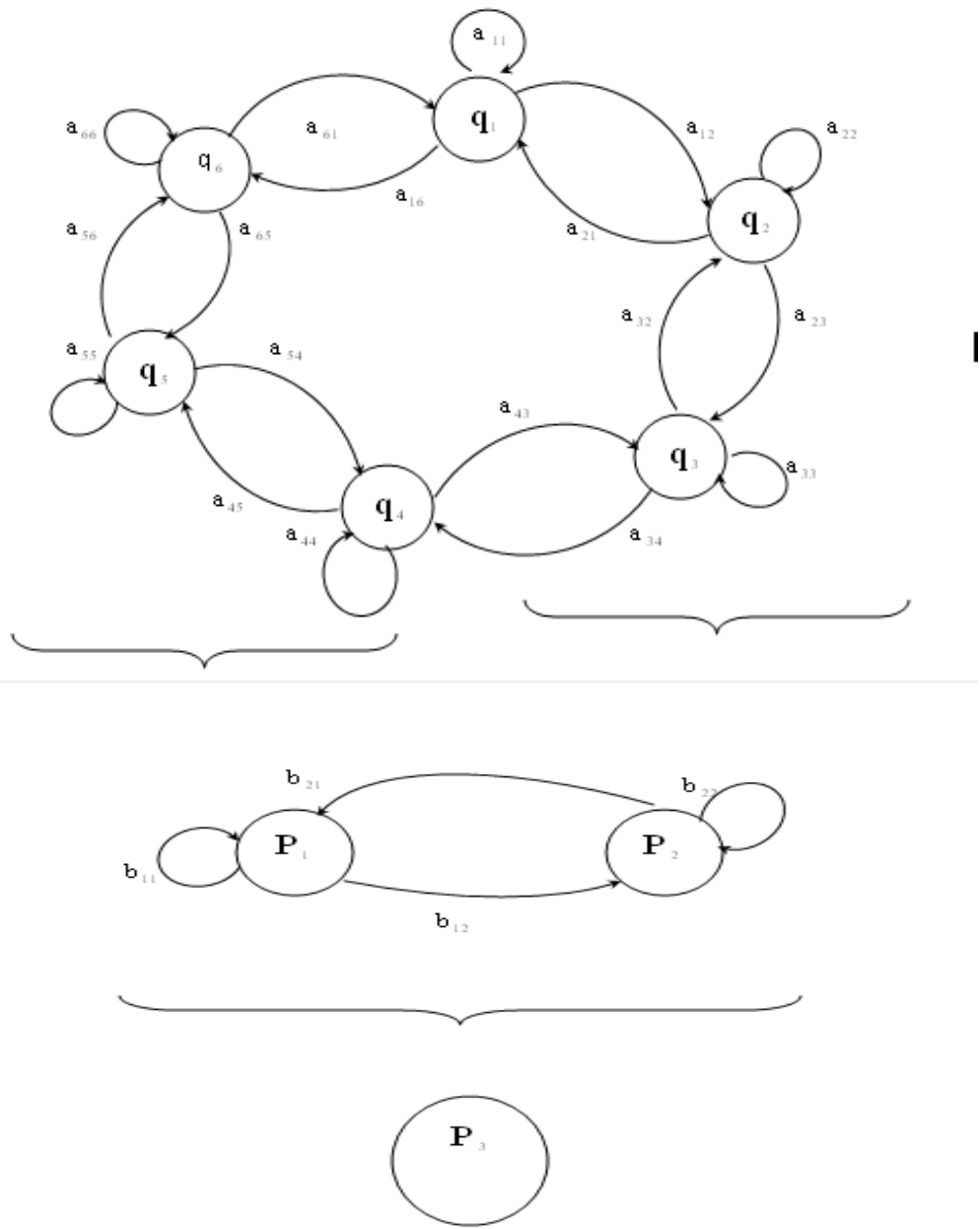

**Figure 2.** Basic structure of CSPHMM1s derived from CHMM1s

### 4.1 Basics of HMM3s

In HMM3s, the underlying state sequence is a third-order Markov chain where the stochastic process is expressed by a 4-D matrix ($a_{ijkw}$). Therefore, the transition probabilities in HMM3s are given as [10],

$$a_{ijkw} = \text{Prob}(q_t = s_w | q_{t-1} = s_k, q_{t-2} = s_j, q_{t-3} = s_i) \quad (4)$$

with the constraints,

$$\sum_{w=1}^{N} a_{ijkw} = 1 \qquad N \geq i, j, k \geq 1$$

The probability of the state sequence, $Q \triangleq q_1, q_2, ..., q_T$, is defined as:

$$\text{Prob}(Q) = \Psi_{q_1} a_{q_1 q_2 q_3} \prod_{t=4}^{T} a_{q_{t-3} q_{t-2} q_{t-1} q_t} \quad (5)$$

where $\Psi_i$ is the probability of a state $s_i$ at time $t = 1$, and $a_{ijk}$ is the probability of the transition from a state $s_i$ to a state $s_k$ at time $t = 3$.

Given a sequence of observed vectors, $O \triangleq O_1, O_2, ..., O_T$, the joint state-output probability is expressed as:

$$\text{Prob}(Q, O | \lambda) = \Psi_{q_1} b_{q_1}(O_1) a_{q_1 q_2 q_3} b_{q_3}(O_3) \prod_{t=4}^{T} a_{q_{t-3} q_{t-2} q_{t-1} q_t} b_{q_t}(O_t) \quad (6)$$

## 4.2 Extended Viterbi and Baum-Welch Algorithms of HMM3s

Based on the probability of the partial alignment ending at a transition $(s_k, s_w)$ at times $(t-1, t)$, the most likely state sequence can be obtained as [10]:

$$\delta_t(j, k, w) \triangleq \text{Prob}(q_1, ..., q_{t-2} = s_j, q_{t-1} = s_k, q_t = s_w, O_1, O_2, ..., O_t | \lambda)$$
$$T \geq t \geq 3, N \geq j, k, w \geq 1 \quad (7)$$

Recursive computation can be calculated as:

$$\delta_t(j, k, w) = \max_{N \geq i \geq 1} \{\delta_{t-1}(i, j, k) \cdot a_{ijkw}\} \cdot b_w(O_t) \quad T \geq t \geq 4, N \geq j, k, w \geq 1 \quad (8)$$

The forward function, $\alpha_t(j, k, w)$, which defines the probability of the partial observation sequence, $O_1, O_2, ..., O_T$, and the transition $(s_j, s_k, s_w)$ among times: $t-2$, $t-1$, and $t$ is defined as:

$$\alpha_t(j,k,w) \triangleq \text{Prob}(O_1,...,O_t, q_{t-2}=s_j, q_{t-1}=s_k, q_t=s_w|\lambda)$$
$$T \geq t \geq 3, N \geq j,k,w \geq 1 \qquad (9)$$

$\alpha_t(j,k,w)$ can be computed from the two transitions: $(s_i,s_j,s_k)$ and $(s_j,s_k,s_w)$ between states $s_i$ and $s_w$ as:

$$\alpha_{t+1}(j,k,w) = \sum_{i=1}^{N} \alpha_t(i,j,k) \cdot a_{ijkw} \cdot b_w(O_{t+1}) \quad T-1 \geq t \geq 3, N \geq j,k,w \geq 1 \qquad (10)$$

The summation of the forward variable can be obtained as,

$$\text{Prob}(O|\lambda_{HMM3s}) = \sum_{w=1}^{N} \sum_{k=1}^{N} \sum_{j=1}^{N} \alpha_T(j,k,w) \qquad (11)$$

The backward function, $\beta_t(i,j,k)$, can be defined as:

$$\beta_t(i,j,k) \triangleq \text{Prob}(O_{t+1},...,O_T | q_{t-2}=s_i, q_{t-1}=s_j, q_t=s_k, \lambda)$$
$$T-1 \geq t \geq 3, N \geq i,j,k \geq 1 \qquad (12)$$

The last equation expresses $\beta_t(i,j,k)$ as the probability of the partial observation sequence from $t+1$ to $T$ given the model $\lambda$ and the transition $(s_i,s_j,s_k)$ among times: $t$-2, $t$-1, and $t$. $\beta_t(i,j,k)$ can be calculated from the two transitions: $(s_j,s_k,s_w)$ and $(s_i,s_j,s_k)$ between states $s_w$ and $s_i$ as:

$$\beta_t(i,j,k) = \sum_{w=1}^{N} \beta_{t+1}(j,k,w) \cdot a_{ijkw} \cdot b_w(O_{t+1}) \quad t = T-1, T-2,...,3, N \geq i,j,k \geq 1 \qquad (13)$$

Readers can obtain more information about third-order hidden Markov models from reference [10].

**4.3 LTRSPHMM3s**

LTRSPHMM3s have been developed from acoustic Third-Order Left-to-Right Hidden Markov Models (LTRHMM3s). As an example of these models, the six first-order acoustic left-to-right hidden Markov states of Fig. 1 are replaced by six third-order acoustic hidden Markov states located in a left-to-right form. The suprasegmental third-order states $p_1$ and $p_2$ are positioned in a left-to-right form. In these models, the suprasegmental state $p_3$ becomes a third-order suprasegmental state.

### 4.4 CSPHMM3s

Within CHMM3s, prosodic and acoustic information can be combined and integrated into CSPHMM3s as given by the following formula,

$$log\ P\left(\lambda^v_{CHMM3s}, \Psi^v_{CSPHMM3s} | O\right) = (1-\alpha).\ log\ P\left(\lambda^v_{CHMM3s} | O\right) \\ + \alpha.\ log\ P\left(\Psi^v_{CSPHMM3s} | O\right) \quad (14)$$

where, $\lambda^v_{CHMM3s}$ is the acoustic third-order circular hidden Markov model of the $v^{th}$ speaker and $\Psi^v_{CSPHMM3s}$ is the suprasegmental third-order circular hidden Markov model of the $v^{th}$ speaker.

As an example of CSPHMM3s, the six first-order acoustic circular hidden Markov states of Fig. 2 are replaced by six third-order acoustic circular hidden Markov states organized in the same shape. $p_1$ and $p_2$ become third-order suprasegmental states placed in a circular form. $p_3$ is a third-order suprasegmental state which is made up of $p_1$ and $p_2$.

We believe that CSPHMM3s are superior models to each of LTRSPHMM1s, LTRSPHMM2s, LTRSPHMM3s, CSPHMM1s, and CSPHMM2s for speaker identification because the characteristics of CSPHMM3s are comprised of the characteristics of both CSPHMMs and SPHMM3s:

1. In SPHMM3s, the state sequence is a third-order suprasegmental chain where the stochastic process is expressed by a 4-D matrix since the state-transition probability at time $t+1$ depends on the states of the suprasegmental chain at times $t$, $t$-1, and $t$-2. Consequently, the stochastic process that is specified by a 4-D matrix gives greater speaker identification performance than that specified by either a 2-D matrix (SPHMM1s) or a 3-D matrix (SPHMM2s).

2. Suprasegmental chain in CSPHMMs is more powerful and more efficient than that in LTRSPHMMs to express the changing statistical characteristics that exist in the actual observations of speech signals. This is because:

   a) The underlying Markov chain in CSPHMMs has no final or absorbing state. Therefore, the corresponding SPHMMs can be trained by as long training sequence as needed. This property does not exist in LTRSPHMMs.

   b) The absorbing state in LTRSPHMMs governs the fact that the rest of a single and long observation sequence provides no further information about earlier states once the underlying Markov chain reaches the absorbing state. In speaker recognition systems, it is true that a Markov chain should be able to revisit the earlier states because the states of SPHMMs reflect the vocal organic configuration of the speaker. Therefore, the vocal organic configuration of the speaker is reflected to states more conveniently using CSPHMMs than using LTRSPHMMs. Therefore, it is inconvenient to utilize LTRSPHMMs having one absorbing state for speaker identification systems.

## 5. Speech Database and Extraction of Features

### 5.1 Collected Speech Database

In this work, CSPHMM3s have been tested on our collected speech database. The database is comprised of eight different sentences captured in each of neutral and shouted talking environments. The eight sentences are:

1) *He works five days a week.*
2) *The sun is shining.*
3) *The weather is fair.*
4) *The students study hard.*
5) *Assistant professors are looking for promotion.*
6) *University of Sharjah.*

7) *Electrical and Computer Engineering Department.*
8) *He has two sons and two daughters.*

Fifty (twenty five male students and twenty five female students) healthy adult native speakers of American English were asked to utter the eight sentences. These speakers were untrained to avoid overstressed expressions. The speakers were separately asked to utter each sentence a number of times in each of neutral and shouted talking environments. The total number of utterances recorded in both talking environments were 5400 ((50 speakers × first 4 sentences × 9 repetitions/sentence in neutral talking environment) + (50 speakers × last 4 sentences × 9 repetitions/sentence × 2 talking environments)).

The captured database was collected in a clean environment by a speech acquisition board using a 16-bit linear coding A/D converter and sampled at a sampling rate of 16 kHz. The database was a wideband 16-bit per sample linear data. A high emphasis filter, $H(z) = 1 - 0.95z^{-1}$, was applied to the collected speech signals. The emphasized speech signals were applied every 5 ms to a 30 ms Hamming window.

**5.2 Extraction of Features**

Mel-Frequency Cepstral Coefficients (static MFCCs) and delta Mel-Frequency Cepstral Coefficients (delta MFCCs) have been used in this work to characterize the phonetic content of speech signals. Such coefficients have been adopted in stressful speech and speaker recognition areas because these coefficients outperform other coefficients in the two areas and because they yield a high-level approximation of a human auditory perception [29], [30], [31].

A 32-dimension feature analysis of both static MFCCs and delta MFCCs (16 static MFCCs and 16 delta MFCCs) has been used in the present work to form the observation vectors in each of LTRSPHMM1s, LTRSPHMM2s, LTRSPHMM3s, CSPHMM1s, CSPHMM2s, and CSPHMM3s. The number of conventional states, $N$, is nine and the number of suprasegmental states is three (each suprasegmental state is comprised of three conventional states) in each suprasegmental model with a continuous mixture observation density has been chosen for each model.

## 6. Speaker Identification Algorithm Based on Each of LTRSPHMM1s, LTRSPHMM2s, LTRSPHMM3s, CSPHMM1s, CSPHMM2s, and CSPHMM3s and the Experiments

The training phase of each of LTRSPHMM1s, LTRSPHMM2s, LTRSPHMM3s, CSPHMM1s, CSPHMM2s, and CSPHMM3s is very alike to the training phase of the conventional LTRHMM1s, LTRHMM2s, LTRHMM3s, CHMM1s, CHMM2s, and CHMM3s, respectively. In the training phase of each of LTRSPHMM1s, LTRSPHMM2s, LTRSPHMM3s, CSPHMM1s, CSPHMM2s, and CSPHMM3s (completely six separate and independent training phases), suprasegmental: first-order left-to-right, second-order left-to-right, third-order left-to-right, first-order circular, second-order circular, and third-order circular models are trained on top of acoustic: first-order left-to-right, second-order left-to-right, third-order left-to-right, first-order circular, second-order circular, and third-order circular models, respectively. In each training phase, the $v^{th}$ speaker model has been derived using the first four sentences of the speech database with 9 repetitions per sentence uttered by the $v^{th}$ speaker in neutral talking environment. The total number of utterances that have been used to derive the $v^{th}$ speaker model in each training phase are 36 (first 4 sentences $\times$ 9 repetitions/sentence).

In the test (identification) phase of each of LTRSPHMM1s, LTRSPHMM2s, LTRSPHMM3s, CSPHMM1s, CSPHMM2s, and CSPHMM3s (totally six separate and independent test phases), each one of the fifty speakers separately uses each one of the last four sentences of the database (text-independent) with 9 repetitions per sentence in each of neutral and shouted talking environments. The total number of utterances that have been used in each evaluation phase per talking environment are 1800 (50 speakers $\times$ last 4 sentences × 9 repetitions/sentence). The probability of generating every utterance per speaker is separately computed based on each of LTRSPHMM1s, LTRSPHMM2s, LTRSPHMM3s, CSPHMM1s, CSPHMM2s, and CSPHMM3s. For each one of these six suprasegmental models, the model with the highest probability is chosen as the output of speaker identification as given in the following formula per talking environment:

$$V^* = \arg \max_{50 \geq v \geq 1} \left\{ P\left(O \mid \lambda_{model}^v, \Psi_{model}^v \right) \right\} \quad (15)$$

where $O$ is the observation vector or sequence that belongs to the unknown speaker, $\lambda_{model}^v$ is the acoustic hidden Markov model (this model can be one of: LTRHMM1s, LTRHMM2s, LTRHMM3s, CHMM1s, CHMM2s, and CHMM3s) of the $v^{th}$ speaker and $\Psi_{model}^v$ is the suprasegmental hidden Markov model (this model can be one of: LTRSPHMM1s, LTRSPHMM2s, LTRSPHMM3s, CSPHMM1s, CSPHMM2s, and CSPHMM3s) of the $v^{th}$ speaker.

## 7. Results and Discussion

This work proposes, implements, and tests novel classifiers called CSPHMM3s for speaker identification in each of neutral and shouted talking environments. In this work, the weighting factor ($\alpha$) has been selected to be equal to 0.5 to avoid biasing towards either acoustic or prosodic model.

To assess the proposed models, speaker identification performance based on such models has been separately compared with that based on each of LTRSPHMM1s, LTRSPHMM2s, LTRSPHMM3s, CSPHMM1s, and CSPHMM2s in each of neutral and shouted talking environments. Text-independent speaker identification performance in each of neutral and shouted talking environments using the collected database based on each of LTRSPHMM1s, LTRSPHMM2s, LTRSPHMM3s, CSPHMM1s, CSPHMM2s, and CSPHMM3s is given in Table 1. This table apparently demonstrates that each model performs almost ideal in neutral talking environment. This is because each acoustic model (LTRHMM1s, LTRHMM2s, LTRHMM3s, CHMM1s, CHMM2s, and CHMM3s) yields very high text-independent speaker identification performance in such a talking environment as given in Table 2. On the other hand, the suprasegmental models perform non-ideally in shouted talking environment as shown in Table 1. This is because each corresponding acoustic model gives low speaker identification performance in this talking environment as shown in Table 2.

Table 1
Text-independent speaker identification performance in each of neutral and shouted talking environments using the collected database based on each of LTRSPHMM1s, LTRSPHMM2s, LTRSPHMM3s, CSPHMM1s, CSPHMM2s, and CSPHMM3s

| Models | Gender | Speaker identification performance (%) | |
|---|---|---|---|
| | | Neutral talking environment | Shouted talking environment |
| LTRSPHMM1s | Male | 96.6 | 73.5 |
| | Female | 96.8 | 75.7 |
| | Average | 96.7 | 74.6 |
| LTRSPHMM2s | Male | 97.5 | 78.9 |
| | Female | 97.5 | 77.9 |
| | Average | 97.5 | 78.4 |
| LTRSPHMM3s | Male | 97.9 | 81.6 |
| | Female | 98.1 | 81.8 |
| | Average | 98.0 | 81.7 |
| CSPHMM1s | Male | 97.4 | 78.3 |
| | Female | 98.4 | 79.1 |
| | Average | 97.9 | 78.7 |
| CSPHMM2s | Male | 98.9 | 82.9 |
| | Female | 98.7 | 83.9 |
| | Average | 98.8 | 83.4 |
| CSPHMM3s | Male | 99.0 | 85.7 |
| | Female | 99.2 | 85.9 |
| | Average | 99.1 | 85.8 |

Table 2
Text-independent speaker identification performance in each of neutral and shouted talking environments using the collected database based on each of LTRHMM1s, LTRHMM2s, LTRHMM3s, CHMM1s, CHMM2s, and CHMM3s

| Models | Gender | Speaker identification performance (%) | |
|---|---|---|---|
| | | Neutral talking environment | Shouted talking environment |
| LTRHMM1s | Male | 90 | 20 |
| | Female | 91 | 22 |
| | Average | 90.5 | 21 |
| LTRHMM2s | Male | 91 | 55 |
|  | Female | 94 | 58 |
|  | Average | 92.5 | 56.5 |
| LTRHMM3s | Male | 93 | 63 |
|  | Female | 95 | 64 |
|  | Average | 94 | 63.5 |
| CHMM1s | Male | 92 | 41 |
|  | Female | 92.4 | 43 |
|  | Average | 92.2 | 42 |
| CHMM2s | Male | 94 | 62 |
|  | Female | 94 | 62.8 |
|  | Average | 94 | 62.4 |
| CHMM3s | Male | 95 | 67.2 |
|  | Female | 95.6 | 68.2 |
|  | Average | 95.3 | 67.7 |

Table 3 summarizes improvement rate of speaker identification performance in shouted talking environment based on CSPHMM3s over that based on each of LTRSPHMM1s, LTRSPHMM2s, LTRSPHMM3s, CSPHMM1s, and CSPHMM2s and number of extra characteristics possessed by CSPHMM3s as compared to each of LTRSPHMM1s, LTRSPHMM2s, LTRSPHMM3s, CSPHMM1s, and CSPHMM2s. It is apparent from this table that as the number of extra characteristics possessed by CSPHMM3s compared to the other models decreases, the improvement rate decreases. This shows the significance of CSPHMM3s and their ability to enhance speaker identification performance in this talking environment compared to the other models. For example, 15.0% (improvement rate of using CSPHMM3s over using LTRSPHMM1s) is higher than 2.9% (improvement rate of using CSPHMM3s over using CSPHMM2s). This is because CSPHMM3s have the advantage of possessing three extra characteristics compared to LTRSPHMM1s: circular hidden Markov models outperform left-to-right hidden Markov models (one extra characteristic) and third-order hidden Markov models are superior to first-order hidden Markov models (two extra characteristics). On the other hand, CSPHMM3s have the advantage of possessing only one extra characteristic compared to CSPHMM2s: third-order hidden Markov models insignificantly lead second-order hidden Markov models.

Table 3
Improvement rate of speaker identification performance in shouted talking environment based on CSPHMM3s over that based on each of LTRSPHMM1s, LTRSPHMM2s, LTRSPHMM3s, CSPHMM1s, and CSPHMM2s and the number of extra characteristics possessed by CSPHMM3s compared to each of LTRSPHMM1s, LTRSPHMM2s, LTRSPHMM3s, CSPHMM1s, and CSPHMM2s

| Models | Improvement rate (%) | Number of extra characteristics |
|---|---|---|
| LTRSPHMM1s | 15.0 | 3 |
| LTRSPHMM2s | 9.4 | 2 |
| LTRSPHMM3s | 5.0 | 1 |
| CSPHMM1s | 9.0 | 2 |
| CSPHMM2s | 2.9 | 1 |

A statistical significance test has been carried out to demonstrate whether speaker identification performance differences (speaker identification performance based on CSPHMM3s and that based on each of LTRSPHMM1s, LTRSPHMM2s, LTRSPHMM3s, CSPHMM1s, and CSPHMM2s in each of neutral and shouted talking environments) are real or simply due to statistical fluctuations. The statistical significance test has been conducted based on the Student's *t* Distribution test as given by the following formula,

$$t_{model\ 1, model\ 2} = \frac{\overline{x}_{model\ 1} - \overline{x}_{model\ 2}}{SD_{pooled}} \quad (16)$$

where $\overline{x}_{model\ 1}$ is the mean of the first sample (model 1) of size *n*, $\overline{x}_{model\ 2}$ is the mean of the second sample (model 2) of the same size, and $SD_{pooled}$ is the pooled standard deviation of the two samples (models) given as,

$$SD_{pooled} = \sqrt{\frac{SD_{model\ 1}^2 + SD_{model\ 2}^2}{2}} \quad (17)$$

where $SD_{model\ 1}$ is the standard deviation of the first sample (model 1) of size *n* and $SD_{model\ 2}$ is the standard deviation of the second sample (model 2) of the same size.

In this work, the calculated *t* values between CSPHMM3s and each of LTRSPHMM1s, LTRSPHMM2s, LTRSPHMM3s, CSPHMM1s, and CSPHMM2s in each of neutral and shouted talking environments using the collected database are tabulated in Table 4. Based on this table, each calculated *t* value in neutral talking environment is smaller than the tabulated critical value $t_{0.05}$ = 1.645 at *0.05* significant level. On the other hand, in shouted talking environment, each calculated *t* value is higher than the tabulated critical value $t_{0.05}$ = 1.645. Hence, CSPHMM3s outperform each of LTRSPHMM1s, LTRSPHMM2s, LTRSPHMM3s, CSPHMM1s, and CSPHMM2s in a shouted talking environment. The reason of this superiority is that CSPHMM3s possess the combined characteristics of each of LTRSPHMM1s, LTRSPHMM2s, LTRSPHMM3s, CSPHMM1s, and CSPHMM2s as was mentioned in Section 4.4. In neutral talking environment, the superiority of CSPHMM3s over each of the other five models becomes minor since the acoustic models: LTRHMM1s, LTRHMM2s, LTRHMM3s, CHMM1s, CHMM2s, and CHMM3s perform well in such a talking environment as given in Table 2.

Table 4
Calculated *t* values between CSPHMM3s and each of LTRSPHMM1s, LTRSPHMM2s, LTRSPHMM3s, CSPHMM1s, and CSPHMM2s in each of neutral and shouted talking environments using the collected database

| | Calculated *t* value | |
|---|---|---|
| $t_{\text{model 1, model 2}}$ | Neutral environment | Shouted environment |
| $t_{\text{CSPHMM3s, LTRSPHMM1s}}$ | 1.197 | 1.932 |
| $t_{\text{CSPHMM3s, LTRSPHMM2s}}$ | 1.335 | 1.864 |
| $t_{\text{CSPHMM3s, LTRSPHMM3s}}$ | 1.492 | 1.801 |
| $t_{\text{CSPHMM3s, CSPHMM1s}}$ | 1.531 | 1.782 |
| $t_{\text{CSPHMM3s, CSPHMM2s}}$ | 1.578 | 1.753 |

Table 5 shows calculated *t* values between each suprasegmental model and its corresponding acoustic model in each of neutral and shouted talking environments using the collected database. This table apparently illustrates that each suprasegmental model is superior to its belonging acoustic model in each talking environment since each calculated *t* value in this table is greater than the tabulated critical value $t_{0.05}$ = 1.645.

Table 5
Calculated *t* values between each suprasegmental model and its corresponding acoustic model in each of neutral and shouted talking environments using the collected database

| | Calculated *t* value | |
|---|---|---|
| $t_{\text{sup. model, acoustic model}}$ | Neutral environment | Shouted environment |
| $t_{\text{LTRSPHMM1s, LTRHMM1s}}$ | 1.712 | 1.857 |
| $t_{\text{LTRSPHMM2s, LTRHMM2s}}$ | 1.789 | 1.892 |
| $t_{\text{LTRSPHMM3s, LTRHMM3s}}$ | 1.826 | 1.955 |
| $t_{\text{CSPHMM1s, CHMM1s}}$ | 1.701 | 1.798 |
| $t_{\text{CSPHMM2s, CHMM2s}}$ | 1.786 | 1.892 |
| $t_{\text{CSPHMM3s, CHMM3s}}$ | 1.894 | 1.986 |

Six more experiments have been independently conducted in this work to assess the achieved speaker identification performance in each of neutral and shouted talking environments based on CSPHMM3s. The six experiments are:

1. Experiment 1: The six classifiers: LTRSPHMM1s, LTRSPHMM2s, LTRSPHMM3s, CSPHMM1s, CSPHMM2s, and CSPHMM3s have been evaluated on a well-known speech database called Speech Under Simulated and Actual Stress (SUSAS). SUSAS database has been originally designed for speech recognition in neutral and stressful talking conditions [12], [13]. In the current work and using this database, isolated words captured at a sampling rate of 8 kHz have been used under each of neutral and angry talking conditions in this experiment. Angry talking condition has been used as an alternative talking condition to shouted talking condition because angry talking condition can not be completely separated from shouted talking condition in our genuine life [6-10]. Thirty different utterances uttered by seven speakers in each of neutral and angry talking conditions have been selected to evaluate the six models. This number of speakers is very limited compared to the number of speakers used in the collected database.

Average speaker identification performance in each of neutral and angry talking conditions using SUSAS database based on each of LTRSPHMM1s, LTRSPHMM2s, LTRSPHMM3s, CSPHMM1s, CSPHMM2s, and CSPHMM3s is exemplified in Fig. 3. This figure evidently shows that the performance based on each model is nearly perfect in neutral talking condition. This figure demonstrates also that speaker identification performance in angry talking condition based on CSPHMM3s is higher than that based on each of the other models. Speaker identification performance based on each model and using SUSAS database is so close to that using the collected database.

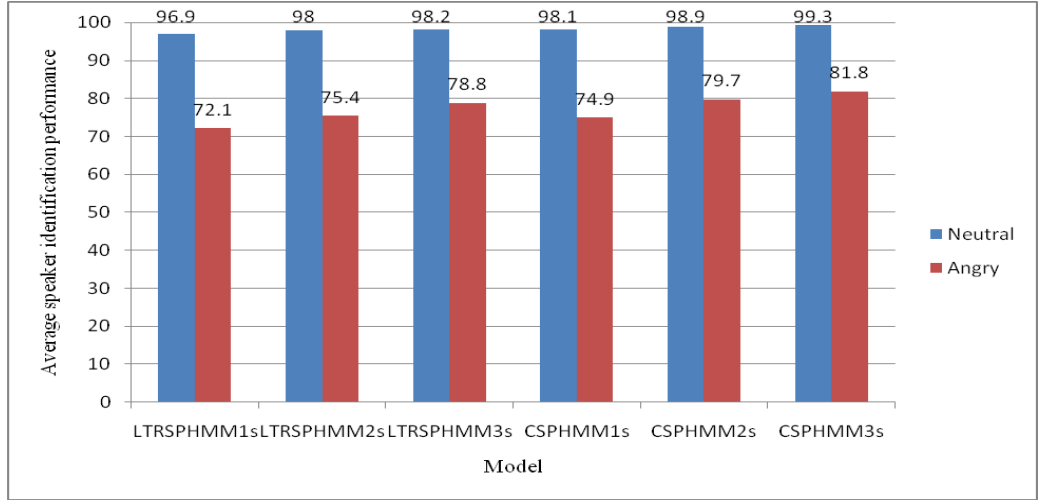

**Figure 3.** Average speaker identification performance in each of neutral and angry talking conditions using SUSAS database based on each of LTRSPHMM1s, LTRSPHMM2s, LTRSPHMM3s, CSPHMM1s, CSPHMM2s, and CSPHMM3s

The calculated *t* values between CSPHMM3s and each of LTRSPHMM1s, LTRSPHMM2s, LTRSPHMM3s, CSPHMM1s, and CSPHMM2s in each of neutral and angry talking conditions using SUSAS database are tabulated in Table 6. This table shows that CSPHMM3s outperform each one of the other five suprasegmental models in angry talking condition (the *t* values are larger than $t_{0.05}$ = 1.645). However, in neutral talking condition, the table illustrates that CSPHMM3s perform almost the same as the other five models (the *t* values are smaller than $t_{0.05}$ = 1.645).

Table 6
Calculated *t* values between CSPHMM3s and each of LTRSPHMM1s, LTRSPHMM2s, LTRSPHMM3s, CSPHMM1s, and CSPHMM2s in each of neutral and angry talking conditions using SUSAS database

|  | Calculated *t* value | |
|---|---|---|
| $t_{\text{model 1, model 2}}$ | Neutral condition | Angry condition |
| $t_{\text{CSPHMM3s, LTRSPHMM1s}}$ | 1.266 | 1.793 |
| $t_{\text{CSPHMM3s, LTRSPHMM2s}}$ | 1.297 | 1.874 |
| $t_{\text{CSPHMM3s, LTRSPHMM3s}}$ | 1.378 | 1.896 |
| $t_{\text{CSPHMM3s, CSPHMM1s}}$ | 1.472 | 1.956 |
| $t_{\text{CSPHMM3s, CSPHMM2s}}$ | 1.501 | 1.987 |

2. Experiment 2: The attained speaker identification performance in shouted/angry talking environment based on CSPHMM3s has been compared with that based on the state-of-the-art models and classifiers using each of the collected and SUSAS databases. Speaker identification performance in shouted/angry talking environment using each of the collected and SUSAS databases based on each of CSPHMM3s, Support Vector Machine (SVM) [32], [33], Genetic Algorithm (GA) [34], [35], and Vector Quantization (VQ) [36], [37] is given in Table 7. This table apparently shows that CSPHMM3s are superior to each of SVM, GA, and VQ for speaker identification in shouted/angry talking environment using each of the collected and SUSAS databases.

Raja and Dandapat [5] reported, based on different classifiers and compensators, speaker identification performance in angry talking environment using SUSAS database [12], [13]. Using MFCCs as the extracted features, their reported speaker identification performance in angry talking environment is 27.0%, 27.0%, 30.2%, 27.0%, 27.0% and 30.2% based, respectively, on Vector Quantization (VQ), Gaussian Mixture Models (GMMs), Speaker and Stress Information based Compensation (SSIC), Compensation by Removal of Stressed Vectors (CRSV), Cepstral Mean Normalization (CMN), and Selection of Compensation by Sress Recognition (SCSR). It is evident that CSPHMM3s yield higher speaker identification performance in angry talking environment using SUSAS database than each of VQ, GMMs, SSIC, CRSV, CMN, and CRSV.

Table 7
Speaker identification performance in shouted/angry talking environment using each of the collected and SUSAS databases based on each of CSPHMM3s, SVM, GA, and VQ

| Classifier | Gender | Speaker identification performance (%) | |
|---|---|---|---|
| | | Using collected database | Using SUSAS database |
| CSPHMM3s | Male | 85.7 | 82.0 |
| | Female | 85.9 | 81.6 |
| | Average | 85.8 | 81.8 |
| SVM | Male | 61.0 | 62.6 |
| | Female | 60.0 | 60.6 |
| | Average | 60.5 | 61.6 |
| GA | Male | 57.0 | 58.5 |
| | Female | 58.0 | 58.1 |
| | Average | 57.5 | 58.3 |
| VQ | Male | 58.0 | 56.2 |
| | Female | 58.0 | 57.0 |
| | Average | 58.0 | 56.6 |

3. Experiment 3: Using our collected database and exploiting CSPHMM3s based on each of our approach, Hadar and Messer approach, and Chatzis approach, the achieved speaker identification performance in shouted talking environment is given in Table 8. This table clearly demonstrates that speaker identification performance using CSPHMM3s based on our approach is greater than that based on Hadar and Messer approach and Chatzis approach by 5.8% and 6.5%, respectively.

Table 8
Speaker identification performance in shouted talking environment using CSPHMM3s based on each of our approach, Hadar and Messer approach, and Chatzis approach using the collected database

| CSPHMM3s based on | Gender | Speaker identification performance in shouted talking environment |
|---|---|---|
| Our approach | Male | 85.7% |
| | Female | 85.9% |
| | Average | 85.8% |
| Hadar and Messer approach | Male | 81.3% |
| | Female | 80.9% |
| | Average | 81.1% |
| Chatzis approach | Male | 80.4% |
| | Female | 80.8% |
| | Average | 80.6% |

4. Experiment 4: CSPHMM3s have been assessed for distinct values of the weighting factor ($\alpha$). Fig. 4 illustrates average speaker identification performance in each of neutral and shouted talking environments using the collected database based on CSPHMM3s for different values of $\alpha$ (0.0, 0.1, 0.2, …, 0.9, 1.0). This figure clearly shows that increasing the value of $\alpha$ in the range [0 - 0.5] has a significant effect on improving speaker identification performance in a shouted talking environment, while increasing the value of $\alpha$ in the range [0.6 – 1.0] has an insignificant influence on enhancing the performance in the same talking environment. This figure demonstrates also that increasing the value of $\alpha$ in the range [0.0 – 1.0] has an insignificant impact on enhancing the performance in a neutral talking environment. In other words, suprasegmental hidden Markov models have more impact than the acoustic models on speaker identification performance in a shouted talking environment.

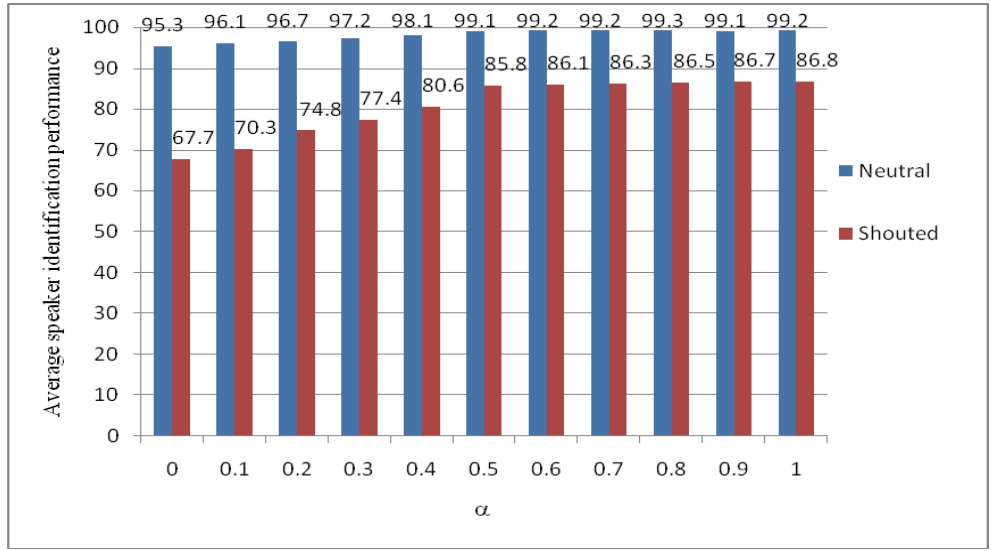

**Figure 4.** Average speaker identification performance in each of neutral and shouted talking environments using the collected database based on CSPHMM3s for different values of $\alpha$

5. Experiment 5: To estimate the standard deviation of speaker identification performance in each of neutral and shouted talking environments based on each of LTRSPHMM1s, LTRSPHMM2s, LTRSPHMM3s, CSPHMM1s, CSPHMM2s, and CSPHMM3s, a statistical cross-validation technique has been conducted in this experiment. Cross-validation technique has been separately performed for each classifier as follows: the whole collected database (5400 utterances per model) has been partitioned at random into five subsets per classifier. Each subset is comprised of 1080 utterances (360 utterances have been used in the training session and the remaining have been used in the test session). Based on these five subsets per classifier, the standard deviation has been computed. The standard deviation values per classifier are summarized in Fig. 5. Cross-validation technique demonstrates, based on this figure, that the calculated standard deviation values are low. As a result, it is evident that speaker identification performance in each of neutral and shouted talking environments based on each classifier and using the five subsets is very similar to that using the entire database.

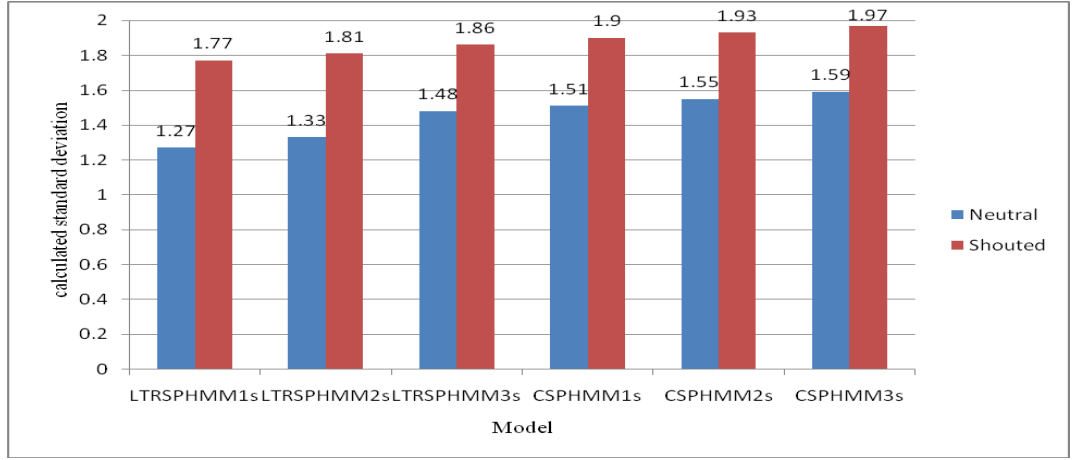

**Figure 5.** Calculated standard deviation values using statistical cross-validation technique in each of neutral and shouted talking environments of the collected database based on each of LTRSPHMM1s, LTRSPHMM2s, LTRSPHMM3s, CSPHMM1s, CSPHMM2s, and CSPHMM3s

6. Experiment 6: An informal subjective evaluation of CSPHMM3s using the collected speech database has been carried out with ten nonprofessional listeners (human judges). A total of 800 utterances (50 speakers × 8 sentences × 2 talking environments) have been used in this evaluation. During the assessment, each listener was separately asked to identify the unknown speaker in each of neutral and shouted talking environments (completely two separate and independent talking environments) for every test utterance. The average speaker identification performance in neutral and shouted talking environments based on the subjective evaluation is 94.8% and 79.4%, respectively. These averages are close to the attained averages based on CSPHMM3s (99.1% and 85.8% in neutral and shouted talking environments, respectively).

Finally, the computational costs and training requirements needed in CSPHMM3s are much grater than those required in each of CSPHMM1s and CSPHMM2s. The required computations of the forward variable, $\alpha_t(j)$, and the backward variable, $\beta_t(j)$, in each of HMM1s, HMM2s, and HMM3s are given in Table 9 where $N \geq j \geq 1$ and $T \geq t \geq 1$ ($N$ is the number of states and $T$ is the utterance length) [10]. It is evident from this table that the required number of computations for each of the forward and backward variables have been dramatically increased in HMM3s compared to each of HMM1s and HMM2s.

Table 9
Required computations of the forward and backward variables in each of HMM1s, HMM2s, and HMM3s

| Models | Required number of computations of the forward variable | Required number of computations of the backward variable |
|---|---|---|
| HMM1s | $N^2 T$ | $N^2 T$ |
| HMM2s | $N^3 T$ | $N^3 T$ |
| HMM3s | $N^4 T$ | $N^4 T$ |

## 8. Concluding Remarks

Novel CSPHMM3s have been proposed, applied, and assessed to improve low text-independent speaker identification performance in shouted/angry talking environment. Some experiments have been separately and independently carried out in this work using different speech databases based on the novel models. This work shows that CSPHMM3s significantly outperform each of LTRSPHMM1s, LTRSPHMM2s, LTRSPHMM3s, CSPHMM1s, and CSPHMM2s in such a talking environment. This is because the characteristics of LTRSPHMM1s, LTRSPHMM2s, LTRSPHMM3s, CSPHMM1s, and CSPHMM2s are all combined and integrated into CSPHMM3s. Also, this work shows that CSPHMM3s slightly perform better than each of CSPHMM1s and CSPHMM2s in neutral talking environment. In addition, this work demonstrates that the suprasegmental models CSPHMM3s are superior to their corresponding acoustic models HMM3s in each of neutral and shouted talking environments. Furthermore, our work exemplifies that CSPHMM3s based on our approach are superior to those based on each of Hadar and Messer approach and Chatzis approach in shouted talking environment. Finally, CSPHMM3s outperform each of SVM, GA, and VQ for speaker identification in shouted talking environment.

There are some limitations in this work. First, using CSPHMM3s for speaker identification nonlinearly increases the computational costs and the needed training requirements compared to using each of CSPHMM1s and CSPHMM2s. This is because HMM3s require on the order of $N^4 T$ operations ($N$ is the number of states and $T$ is the utterance length), compared to $N^2 T$ and $N^3 T$ operations in HMM1s and HMM2s, respectively. Therefore, it is required much more memory space in CSPHMM3s than that in each of CSPHMM1s and CSPHMM2s. Second, the number of speakers accessible in SUSAS database is inadequate. Third, all the available 7 speakers in SUSAS database are of the same gender (male). Finally, speaker identification performance in shouted/angry talking environment based on CSPHMM3s is imperfect. A comprehensive study and investigation are underway to further improve speaker identification performance in such a talking environment.